# Anisotropic complex refractive indices of atomically thin materials: determination of the optical constants of few-layer black phosphorus

**Aaron M. Ross** [1], **Giuseppe M. Paternò** [2], **Stefano Dal Conte** [1], **Francesco Scotognella** [1,2]* and **Eugenio Cinquanta** [3]

[1]  Dipartimento di Fisica, Politecnico di Milano, Piazza Leonardo da Vinci 32, 20133 Milano, Italy
[2]  Center for Nano Science and Technology@PoliMi, Istituto Italiano di Tecnologia (IIT), Via Giovanni Pascoli, 70/3, 20133, Milan, Italy
[3]  Istituto di Fotonica e Nanotecnologie, Consiglio Nazionale delle Ricerche, Milano, Italy
*  Correspondence: francesco.scotognella@polimi.it; Tel.: +39-02-2399-6056

**Abstract:** In this work we briefly review the studies of the optical constants of monolayer transition metal dichalcogenides and few layer black phosphorus, with particular emphasis to the complex dielectric function and refractive index. Specifically, we give an estimate of the complex index of refraction of phosphorene and few-layer black phosphorus. We extracted the complex index of refraction of this material from differential reflectance data reported in literature by employing a constrained Kramers-Kronig analysis. Finally, we studied the linear optical response of multilayer systems embedding phosphorene by using the transfer matrix method.

## 1. Introduction

Two-dimensional materials emerged as a promising option for the development of nanoelectronics and optoelectronics devices, due to their peculiar electronic and optical properties. In this respect, transition metal dichalcogenides (TMDs), when thinned to single atomic layer, show an indirect-direct band gap transition, inequivalent valleys in the Brillouin zone and non-trivial topological order [1–3]. In the vast family of two-dimensional materials, elemental ones represent a niche, due to their high reactivity at ambient conditions that makes their characterization and exploitation challenging [4,5]. Among them, phosphorene, *i.e.* monolayer black phosphorous (BP), is the most promising material for optoelectronic and photonic application as its electronic band gap lies in a spectral region between the ones of TMDs and the one of graphene [6]. For an effective employment of these materials in photonic and optoelectronic devices, it is very important to determine their dielectric response.

Among the monolayers of transition metal dichalcogenides, $MoS_2$ has been the first one on which optical measurements have been performed. In fact, in 2013 and 2014 the complex dielectric function of $MoS_2$ has been measured with spectroscopic ellipsometry by Shen et al. [7], Yim et al. [8] and Li et al. [9]. The complex dielectric functions of monolayers $MoS_2$, $WS_2$, $MoSe_2$, and $WSe_2$ have been measured by Li et al. employing reflectance contrast spectroscopy [10]. In 2019, the monolayers, together with the bilayers and the trilayers, of the four above mentioned chalcogenides have been measured by Hsu et al [11]. In 2020, Ermolaev et al. [12] have measured the complex dielectric function of $MoS_2$ in a broad range of wavelengths, i.e. 290 nm – 3300 nm. Ermolaev et al. [13] have also measured the complex index of refraction of $WS_2$ in the range 375 – 1700 nm.

Additionally, a number of studies have investigated the anisotropic components of TMD and phosphorous few-layer systems via incorporation of dielectric and plasmonic nanostructures [14–21]. In one study, the coupling of electromagnetic fields with the out-of-plane component of the $MoS_2$ dielectric function was possible due to the peculiar morphology of CVD grown few-layers $MoS_2$ on rippled substrate, hence promoting 1D nanostructures as a promising path for the full exploitation of the few layer $MoS_2$ dielectric response [19]. Similarly, a top-down approach revealed to be effective for the strain-induced modification of the electronic band structure of atomically thin $MoS_2$ and black phosphorous, paving the way for the on-demand tuning of their optical properties [20,21] One study investigated exfoliated $MoS_2$ onto a lithographically-defined SiO2 nanocone substrate, demonstrating a deterministic red-shift of the A exciton PL due to elastic strain [14]. Another study utilized spatially and time-resolved PL diffusion measurements of WSe₂ exfoliated onto a 1.5 um diameter SiO₂ pillar, resulting in a similar red-shift and demonstration of exciton diffusion towards high tensile strain regions; that work was a proof of concept for engineerable excitonic diffusion that may lead to a new generation of opto-excitonic devices [15].

Strain engineering may also be used to enable new radiative pathways for optical excitation of selection-rule-forbidden TMD dark excitons under normal incidence conditions; these dark and "grey" excitons possess much longer radiative lifetimes (>100 ps) [16]. In the absence of intentional strain, one study utilized a high numerical aperture objective lens (NA = 0.82) to characterize the dark exciton in WSe₂: even for normal incidence, a Gaussian beam with a tight focus has a non-zero electric field component along the beam propagation direction [16]. However, optical excitation of the dark exciton in WSe₂ was more easily enabled by coupling to the surface plasmon polariton of Ag [17]. Another study incorporated an hBN-encapsulated WSe2 monolayer into a plasmonic modulator device, requiring precise knowledge of the TMD, Au, and hBN complex indices of refraction [18]. It thus emerges how the development of simple and robust tools for the careful analysis of the dispersion of the complex refractive index is beneficial for an appropriate engineering of devices that include two-dimensional materials.

In contrast, phosphorene (single layer BP) and few-layer BP, have not been studied with the same fervor of their TMD counterparts, likely due to their relative sensitivity to degradation. One significant distinction between TMDs and BP is that the band structure of BP exhibits considerable anisotropy: that is, the effective electron mass along the armchair and zig-zag directions is a factor of five times larger in the former direction [22]. This anisotropy is manifest in polarization-sensitive absorption and PL measurements [22–25], as well as spatially-resolved excitonic diffusion measurements [22]. The relatively fragile phosphorene system holds considerable promise due to the ease of PL/absorption tunability by layer number tuning, which can shift the resonance energies between 0.32 and 1.7 eV [22].

We focus on estimating the complex refractive index of phosphorene, which shows a morphological in-plane anisotropy resulting in an anisotropic optical response [23]. We retrieve the refractive index along armchair (AC) and zig-zag (ZZ) direction by exploiting Constrained Kramers-Kronig analysis (CKKA) and fit experimental contrast reflectivity with a model obtained by means of the Transfer Matrix Method (TMM).

## 2. Materials and Methods

*Methodology for extraction of the complex index of refraction of few-layer BP from reflectance contrast data*: The linear optical properties of thin films are commonly measured via static absorption or reflectivity contrast measurements. Experimental techniques include normal incidence reflectivity contrast (dR/R) over a wide plane of view, spatially-resolved confocal dR/R in the case of exfoliated small-area TMDs, few-layer BP, graphene, and graphene oxide, and simultaneous absolute reflectivity and transmission measurements for determination of static absorption [10,11,23,25–28] Although these methods reveal the static absorption $A$ and reflectivity dR/R and transmission dT/T contrasts of the thin films relative to the linear substrate response, the determination of the complex index of refraction $\tilde{n} = n + ik$, or equivalently the complex dielectric function $\varepsilon = \varepsilon_r + i\varepsilon_i$, where $n = \sqrt{\varepsilon}$, typically requires more sophisticated optical characterization methods such as ellipsometry [8,29–31].

In this section, we extract the complex index of refraction of hBN-encapsulated few-layer BP on a sapphire substrate from confocal reflectivity contrast data taken from Reference [23]. Specifically, the layered sample structure is as follows: air → 15 nm hBN → few-layer BP → $\alpha$-Al₂O₃ (Sapphire); BP thicknesses range nominally from 0.5 nm (1L) to 2.5 nm (5L), and bulk BP is 100 nm thick. We assume that the exfoliated hBN layer is crystalline, rather than isotropic as for the case of cubic BN or polycrystalline hBN [31–33] with an in-plane dielectric constant given by $\varepsilon_{hBN,\parallel} = n_{hBN,\parallel}^2 = 1 + \frac{3.336 \lambda^2}{\lambda^2 - 26322}$, where the wavelength $\lambda$ is given in nanometers [31]. Additionally, we also use a Sellmeier equation to describe the $\alpha$-Al₂O₃ substrate index of refraction [34].

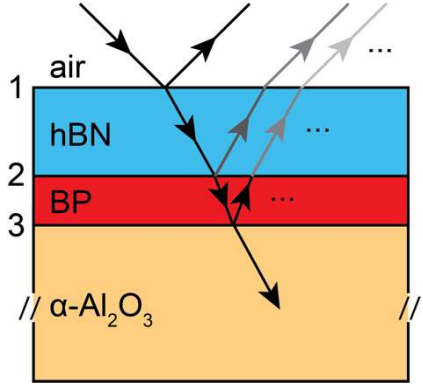

Figure 1: Schematic diagram of multi-layer system under study: air, 15 nm of hBN, 1-5 layers of BP, and a thick substrate of $\alpha$-Al$_2$O$_3$ [23]. Input and output wave are indicated directionally by arrows, where only a few reflected waves from each interface are indicated. The incident angle is shown as oblique for clarity (normal incidence in experiment).

One commonly used method for determination of the complex linear material response functions is the Kramers-Kronig analysis (KKA): it is well-known that the real and imaginary parts of the dielectric function are related to one another via the Kramers-Kronig (KK) relations, which impose causality conditions on the response functions [35–37]. However, the integrals involved in these relations are taken over infinite limits; practical implementation requires the acquisition of data over a wide energy range, and may be limited to materials which exhibit an optical response only over a narrow wavelength range.

We utilize an alternative method called the *constrained* Kramers-Kronig analysis (CKKA), which was introduced by Kuzmenko to address some of the problems associated with KKA [35]. This method takes advantage of the fact that basis functions such as the complex Lorentzian can be chosen that individually satisfy the KK relations. Such a function is given by

$$\varepsilon_k = \frac{\omega_{p,k}^2}{\omega_{0,k}^2 - \omega^2 + i\omega\gamma_k}$$

where $\omega_{p,k}$ is the plasma frequency, $\omega_{0,k}$ is the oscillator frequency, and $\gamma_k$ is the linewidth. Since linear combinations of such basis functions also satisfy the KK relations, the method introduces a mesh in energy space of a large number of oscillators $N_{osc} \sim N_{data}$; the oscillator frequencies are fixed, and the linewidth is chosen as the spacing between oscillator energies. Thus, the only remaining parameters left to minimize are the oscillator strengths (or plasma frequencies): hence, the *constrained* KK analysis. To further optimize the CKKA fitting procedure, we chose functions that are similar to the complex Lorentzian of Eq. 1 and also satisfy the KK relations, but are more locally-weighted, avoiding the contributions of the tails of the Lorentzian distributions far away from the central oscillator energy. One such function is a triangular function for $\varepsilon_i$, with a corresponding analytically integrable $\varepsilon_r$ [35,36].

We note that although this method can easily reproduce reflectivity contrast data when a sufficiently large oscillator mesh is used, it is sensitive to a lack of information about the optical response outside of the experimental acquisition range [10,35,36]. This problem is not entirely unexpected, considering that it is fundamentally a KK-type analysis method. Often, high-frequency information about a given optically active system is known via X-ray photoelectron emission spectroscopy or UV absorption experiments; additionally, low-frequency information is often revealed via Fourier-transform IR (FTIR) and Raman scattering experiments [5,38–40] However, such information is currently lacking for few-layer BP, although bulk BP is well-characterized. Although similar calculations using the CKKA method have extracted the complex index of refraction for TMD monolayer and few-layer systems assuming a bulk-like response at high energies [10], no such assumptions are made here for the few-layer BP fitting procedure.

For a given complex index of refraction, the optical response of a multi-layered system is then determined by the transfer matrix method (TMM) [41]. It is noted that perturbative "linearized" methods have also been developed for the treatment of TMD monolayers, few-layer BP, and graphene [10,23,28] which treat the optical response of the thin film using the sheet conductivity $\sigma^d = \sigma d = -id\omega(\varepsilon - 1)$. For basic layered systems such as air → monolayer TMD → $SiO_2$, relatively simple forms for the reflectivity contrast dR/R can be derived [27]. However, for the system under study in this section (air → hBN → BP → $\alpha$-$Al_2O_3$), with no assumptions being made about reflection coefficients between interfaces, no tractable analytical form for dR/R is derived; note that this stands in contrast to previous studies [23] which assume that the BP → $\alpha$-$Al_2O_3$ interface does not contribute to the overall optical response of the layered system.

We use the standard TMM [41] to relate the input and reflected fields at the air → hBN interface (1) to the transmitted fields at the BP → $\alpha$-$Al_2O_3$ interface (3). Specifically, for normal incidence fields, the electric and magnetic fields, which are implicit sums of all fields (incident, reflected, transmitted) are given by

$$\begin{bmatrix} E_1 \\ H_1 \end{bmatrix} = \widehat{M}_{1-2}\widehat{M}_{2-3}\begin{bmatrix} E_3 \\ H_3 \end{bmatrix}, \widehat{M}_{i-j} = \begin{bmatrix} \cos\phi_{ij} & \frac{i}{n_{ij}}\sin\phi_{ij} \\ in_{ij}\sin\phi_{ij} & \cos\phi_{ij} \end{bmatrix}, \phi_{ij} = \frac{2\pi n_0 n_{ij} d_{ij} E}{hc}$$

which can then be related to the field reflection coefficient by the following relation

$$r = \frac{E_r}{E_i} = \frac{n_0 m_{11} + n_0 n_s m_{12} - m_{21} - n_s m_{22}}{n_0 m_{11} + n_0 n_s m_{12} + m_{21} + n_s m_{22}}$$

where $m_{ij}$ is the *ij*-th element of the product of the two matrices in Eq. 2, $n_0, n_s$ are the indices of refraction of air and the back substrate ($\alpha$-$Al_2O_3$), $hc$ is the product of the speed of light and Planck's constant, $E = hc/\lambda$ is energy, $n_{ij}$ is the index of refraction for the material layer between the *i* and *j* interfaces, and $d_{ij}$ is the thickness between the *i* and *j* interfaces (Figure 1). Finally, the reflectance contrast is given by $dR/R = \left(|r_{ref}|^2 - |r_{sample}|^2\right)/|r_{ref}|^2$, where $r_{ref}$ is the field reflection coefficient for the reference air → hBN → $\alpha$-$Al_2O_3$, which is determined by setting $\widehat{M}_{2-3} = \hat{1}$. This procedure makes no perturbative assumptions about the index of refraction of BP, and allows for generalization to thicker layers such as the bulk BP case.

## 3. Results

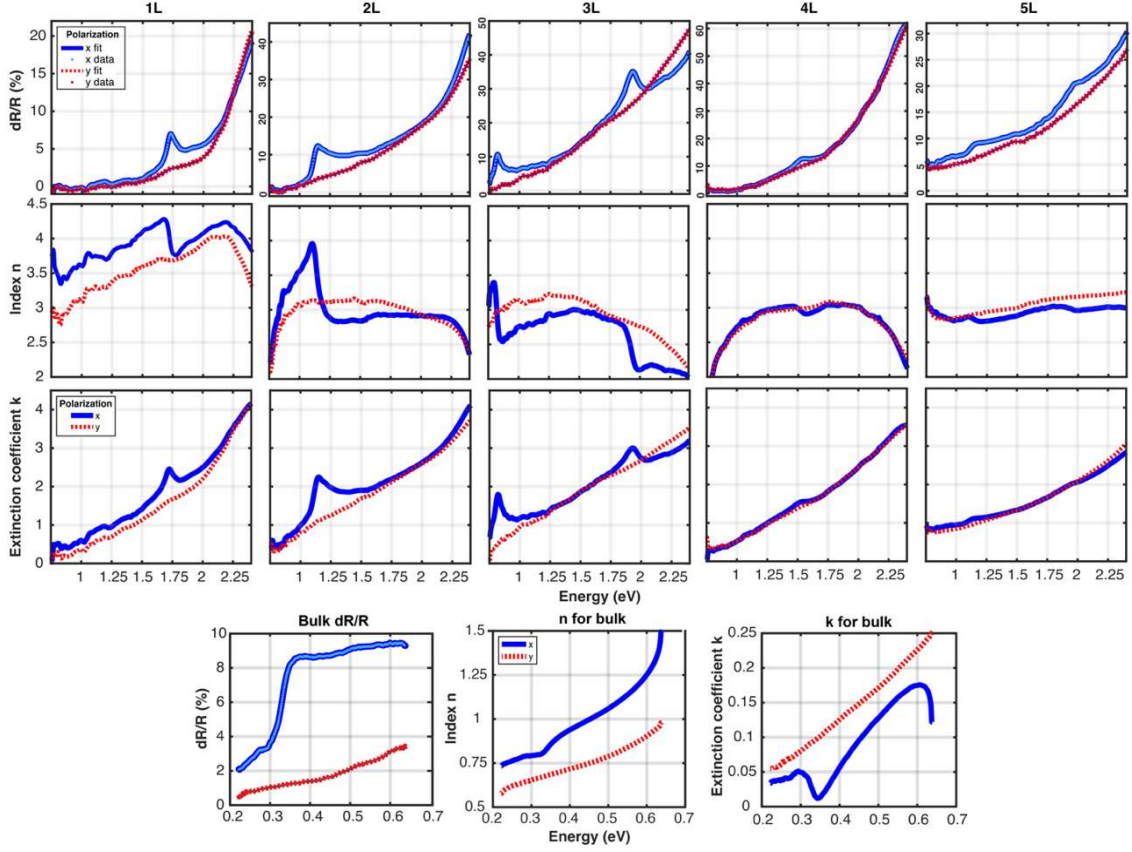

**Figure 2.** Complex index of refraction: results of fitting reflectance contrast dR/R data for 1-5 L and bulk BP, taken from Li. et. al. [23] using the CKKA method. First row: polarization-selective reflectance contrast dR/R ($1 - R_{sample}/R_{ref}$) data, with CKKA + TMM fits. Second row: Real index of refraction n extracted using CKKA + TMM method. Third row: Extinction coefficient k using CKKA + TMM method. Fourth row: dR/R, n, k, for bulk (100 nm) BP.

*Complex index of refraction for few-layer and bulk BP*: The complex index of refraction for few-layer BP is then determined by fitting the reflectance contrast data [23] using non-linear least squares fitting of the weights (plasma frequencies) parameters in the CKKA method with triangular basis functions; this method is referred to as CKKA+TMM [35,36]. The dR/R data was extracted from the literature using a plot digitizer with high accuracy [42]. The resulting dR/R fits, real indices of refraction *n*, and extinction coefficients *k*, for both x- and y-polarized excitation for 1-5 layers and bulk BP are displayed in Figure 2. We note the high quality of the dR/R fits (Figure 1, top row): even small variations in dR/R are easily reproduced with CKKA+TMM: a high-resolution mesh of 300 triangular oscillators was chosen to fit these data. We argue that the *constrained* KK method utilized here provides credibility to the resulting complex indices of refraction, since the dispersive component (*n*, or $\varepsilon_r$) is connected directly via the KK relations to the absorptive component (*k*, or $\varepsilon_i$), which for atomically thin films is the major contribution to the dR/R measurement [10,27].

As it has been reported previously [23], the reflectance contrast data for 1-5L BP (taken at 77 K) shows either 1 or 2 peaks within the experimental acquisition range of 0.75-2.4 eV, corresponding to both the intralayer exciton at lower energies and a blue-shifted exciton that arises due to interlayer effects similar to quantum wells [22]. We note that x and y-polarized excitation here corresponds to polarization aligned along the AC and ZZ directions of the puckered few-layer BP, respectively. This puckering effect leads to significant in-plane anisotropy, resulting in a band structure for which the effective electron mass along the AC direction is five times larger than the ZZ direction [22].

As revealed both in the dR/R data, as well as in our extracted complex indices of refraction, sharp excitonic features are observed only in the AC (x) case in dR/R and *k*, with strongly dispersive lineshapes in *n* imposed on a relatively flat background which varies from 2-4. In the case of ZZ excitation, both the index *n* and extinction coefficient *k* are relatively featureless: the index is relatively flat between 2.5-3 from 1-2.2 eV in the cases of 2-5 L, while the extinction coefficient *k* increases from 0 to 4 between these energy ranges. In fact, for both *n* and *k,* the results for ZZ excitation seem to constitute the background features in the AC case, indicating the presence of shared higher energy resonances which are isotropic. Thus, by examining the extinction coefficient *k* directly, which corresponds to absorption, we confirm previous statements in the literature that there exists a broad and increasing background absorption in few-layer BP towards higher energy, on which the strong and narrow excitonic resonances are super-imposed.

The CKKA+TMM analysis used here is more general than the linearized models often used for the treatment of static absorption by atomically thin materials such as TMDs and few-layer BP [10,23]. Indeed, this model is used here to extract quantitatively meaningful results for the complex index of refraction, without resorting to approximations such as small optical path length for the encapsulation layer, minimal absorption in the BP layers, or the neglect of the substrate reflection at the BP → $\alpha$-Al$_2$O$_3$ interface.

This generality allows us to also fit the 100 nm bulk BP sample (Figure 1, fourth row). The dR/R data is no longer easily directly correlated with the absorption coefficient *k*, due to the non-negligible thickness of bulk BP, as well as the interference effects with the hBN encapsulating layer. High quality fits of the dR/R data reveal again large anisotropy resulting in birefringence of bulk BP, with a modulation of the index of refraction around 0.32 eV in the AC case, which is not observed in the ZZ excitation case. For the extinction coefficient *k*, in the AC case, the absorption shows an onset around the band-edge at 0.32 eV, but also shows absorption at lower energies likely due to the Drude response from free carrier excitation [22]. Surprisingly, for the ZZ excitation case, the absorption also increases steadily towards higher energies, without any dispersion modulation around the band edge.

We also note that in this energy range the extracted index of refraction for bulk BP is below 1. Sub-unity indices of refraction are commonly reported in noble metal materials such as Ag and Au, where there is an interplay between interband absorption and the plasmonic response [43–45] as well as in Al thin films at wavelengths shorter than 660 nm [46]. Indeed, a zero-crossing of the real part of the dielectric function $\varepsilon_r$ is required to satisfy the conditions for a localized surface plasmon resonance (LSPR) [47]: the plasmonic nature of BP is currently under investigation, due its novel features such as hyperbolic plasmonics arising from the large ratio between the AC and ZZ direction effective electron mass [22].

*Systematics*: In this section we remark on the sensitivity of the CKKA+TMM analysis for extraction of the complex indices of refraction to variations in parameters such as substrate and encapsulation layer refractive indices, as well as fixed fitting parameters. Previous reports involving the utilization of the CKKA method have commented on the need for incorporating information about low and high energy extremes of the dielectric functions outside of the experimental acquisition range. For instance, Li et. al. [10] assumed during their studies of TMD monolayers that at high energies (> 3 eV), the bulk optical properties contributed also to the monolayer optical response: absorption resonances were included up to 30 eV. No such assumptions are made in our determination of *n,k* for few-layer BP or the bulk BP.

However, in Figure 3, a convergence of the fitting results for 1L-x BP is shown as a function of the *buffer energy* that is included in the fits. More specifically, the experimental data were collected from 0.75-2.4 eV; additionally, we examined the convergence of the fit solution as a function of how far the oscillator energy mesh extended beyond these extremal values, ie. $E_{mesh} \in [E_{data,i} - E_{buf}, E_{data,f} + E_{buf}]$. Although the imaginary part of the CKKA basis triangular dielectric function is well-localized in energy, the real part exhibits long tails away from the oscillator center: thus, by extending our oscillator energy range, we can simulate how these low and high energy contributions increase the quality of fit. This buffer energy was tested from 0 to 0.3 eV: it is shown (Figure 3, center, inset) that increasing the buffer energy from 0 to 0.1 eV reduces the sum of the residuals by nearly two orders of magnitude, with generally no enhancement beyond 0.1 eV. This convergence can be seen in the results for *n* and *k*. However, it is noted that at energies greater than 2.4 eV that the real index of refraction *n* varies considerably for different buffer energies; random variations are also observed in this energy range for slight variations in the number of energy mesh points (not shown). We believe that this lack of convergence arises because of the lack of constraints on the dielectric function we have imposed on energies higher than 2.5 eV; this aspect may be potentially improved by using the higher energy absorption data for bulk BP, in a similar manner as done with TMD monolayers.

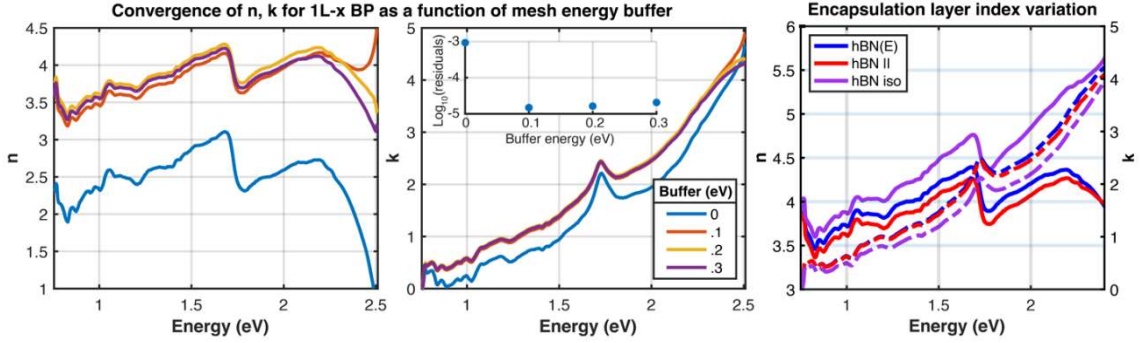

**Figure 3.** Systematic variation in CKKA+TMM parameters. Left and center plots: demonstration of convergence of CKKA+TMM fits for x-polarized excitation of 1L BP as a function of mesh energy buffer. This buffer is defined as the energy range extending beyond the experimental data acquisition range used in the CKKA+TMM fit. Left: real index of refraction n as a function of buffer energy in eV (see legend in plot on right). Center: Extinction coefficient k as a function of buffer energy. Inset on center: Log$_{10}$ of the residuals of the dR/R fits as a function of buffer energy. Residuals are defined here as $\sum |x_{fit,j} - x_{data,j}|^2$. Right: variation of the index of refraction for the 15 nm hBN encapsulating layer. hBN(E), hBN ∥, and hBN iso correspond to using the Sellmeier equation form of the index of refraction for in-plane hBN , a constant value of $n_{hBN}$ = 2.108 [31], and a constant value $n_{hBN}$ = 1.86 [32], respectively. Solid lines correspond to n of BP, and dashed lines correspond to k of BP.

The dependence of the results for 1L-x BP were also studied as a function of the hBN refractive index. It is known that exfoliated hBN is birefringent, but estimates of the for in-plane and out-of-plane indices of refraction have varied significantly between reports [31–33]. For the results displayed in Figure 2, as well as Figure 3, a Sellmeier equation is used that describes the index of refraction of the in-plane index, which ranges from 2.25 to 2.1 from 400-1600 nm [31]. The fitting procedure was also performed for a constant index $n$ = 2.108, or the average of the in-plane index, and $n$ = 1.86, a commonly used value for isotropic BN [31–33]. Using the constant in-plane value yields a generally lower value than the Sellmeier case, and using the isotropic value leads to much larger estimates of $n$; the variations in $k$ are not as significant. We believe that the use of the functional Sellmeier equation form is appropriate: the measurement resulting in that form utilized confocal oblique incidence ellipsometry to capture both the in-plane and out-of-plane indices from 400-1600 nm [31].

## 4. Discussion

For the estimate of the index of refraction of black phosphorus, we have demonstrated that the constrained KK analysis combined with a general transfer matrix method (CKKA+TMM) that takes into account finite layer optical path lengths can be used to extract the entire complex index of refraction of 1-5 L BP and bulk BP from reflectance contrast data. It is confirmed that the excitonic peaks correspond to peaks in the extinction coefficient $k$, and that a significant absorptive background that increases towards higher energies exists. In contrast, the real index of refraction is relatively flat with superimposed dispersive features located at the excitonic resonances. The determination of the full complex index of refraction $\tilde{n}$ of few-layer BP is significant: these results may be incorporated into the development of novel optical devices such as responsive photonic crystals and distributed Bragg reflectors (DBRs) [48–51], plasmonic nanostructures [52], and on-chip waveguides, all of which require knowledge of both the dispersive and absorptive optical properties. We suggest further improvements to this extractive method that would involve the experimental acquisition of both reflectance and transmission contrast simultaneously, as well as constraining the low and high energy ranges in the CKKA+TMM analysis with optical properties deduced from FTIR and UV absorption.

**Author Contributions:** Conceptualization, A.M.R., G.M.P., S.D.C., F.S. and E.C.; methodology, validation and formal analysis A.M.R., F.S. and E.C.; writing—original draft preparation, A.M.R. and F.S; writing—review and editing, A.M.R., G.M.P., S.D.C., F.S. and E.C. All authors have read and agreed to the published version of the manuscript.

**Funding:** This project has received funding from the European Research Council (ERC) under the European Union's Horizon 2020 research and innovation programme (grant agreement No. [816313]).

**Conflicts of Interest:** The authors declare no conflict of interest

## References

1. Mak, K.F.; Lee, C.; Hone, J.; Shan, J.; Heinz, T.F. Atomically Thin MoS2: A New Direct-Gap Semiconductor. *Phys. Rev. Lett.* **2010**, *105*, 136805, doi:10.1103/PhysRevLett.105.136805.

2. Zeng, H.; Dai, J.; Yao, W.; Xiao, D.; Cui, X. Valley polarization in MoS2 monolayers by optical pumping. *Nat. Nanotechnol.* **2012**, *7*, 490–493, doi:10.1038/nnano.2012.95.

3. Huang, L.; McCormick, T.M.; Ochi, M.; Zhao, Z.; Suzuki, M.-T.; Arita, R.; Wu, Y.; Mou, D.; Cao, H.; Yan, J.; et al. Spectroscopic evidence for a type II Weyl semimetallic state in MoTe 2. *Nat. Mater.* **2016**, *15*, 1155–1160, doi:10.1038/nmat4685.

4. Tao, L.; Cinquanta, E.; Chiappe, D.; Grazianetti, C.; Fanciulli, M.; Dubey, M.; Molle, A.; Akinwande, D. Silicene field-effect transistors operating at room temperature. *Nat. Nanotechnol.* **2015**, *10*, 227–231, doi:10.1038/nnano.2014.325.

5. Wood, J.D.; Wells, S.A.; Jariwala, D.; Chen, K.-S.; Cho, E.; Sangwan, V.K.; Liu, X.; Lauhon, L.J.; Marks, T.J.; Hersam, M.C. Effective Passivation of Exfoliated Black Phosphorus Transistors against Ambient Degradation. *Nano Lett.* **2014**, *14*, 6964–6970, doi:10.1021/nl5032293.

6. Churchill, H.O.H.; Jarillo-Herrero, P. Phosphorus joins the family. *Nat. Nanotechnol.* **2014**, *9*, 330–331, doi:10.1038/nnano.2014.85.

7. Shen, C.-C.; Hsu, Y.-T.; Li, L.-J.; Liu, H.-L. Charge Dynamics and Electronic Structures of Monolayer MoS2 Films Grown by Chemical Vapor Deposition. *Appl. Phys. Express* **2013**, *6*, 125801, doi:10.7567/APEX.6.125801.

8. Yim, C.; O'Brien, M.; McEvoy, N.; Winters, S.; Mirza, I.; Lunney, J.G.; Duesberg, G.S. Investigation of the optical properties of MoS 2 thin films using spectroscopic ellipsometry. *Appl. Phys. Lett.* **2014**, *104*, 103114, doi:10.1063/1.4868108.

9. Li, W.; Birdwell, A.G.; Amani, M.; Burke, R.A.; Ling, X.; Lee, Y.-H.; Liang, X.; Peng, L.; Richter, C.A.; Kong, J.; et al. Broadband optical properties of large-area monolayer CVD molybdenum disulfide. *Phys. Rev. B* **2014**, *90*, 195434, doi:10.1103/PhysRevB.90.195434.

10. Li, Y.; Chernikov, A.; Zhang, X.; Rigosi, A.; Hill, H.M.; Van Der Zande, A.M.; Chenet, D.A.; Shih, E.M.; Hone, J.; Heinz, T.F. Measurement of the optical dielectric function of monolayer transition-metal dichalcogenides: MoS2, MoSe2, WS2, and WSe2. *Phys. Rev. B* **2014**, *90*, 1–6, doi:10.1103/PhysRevB.90.205422.

11. Hsu, C.; Frisenda, R.; Schmidt, R.; Arora, A.; Vasconcellos, S.M.; Bratschitsch, R.; der Zant, H.S.J.; Castellanos-Gomez, A. Thickness-Dependent Refractive Index of 1L, 2L, and 3L MoS2 , MoSe2 , WS2 , and WSe2. *Adv. Opt. Mater.* **2019**, 1900239, doi:10.1002/adom.201900239.

12. Ermolaev, G.A.; Stebunov, Y. V; Vyshnevyy, A.A.; Tatarkin, D.E.; Yakubovsky, D.I.; Novikov, S.M.; Baranov,

D.G.; Shegai, T.; Nikitin, A.Y.; Arsenin, A. V; et al. Broadband optical properties of monolayer and bulk MoS 2. *npj 2D Mater. Appl.* **2020**, *4*, 1–6, doi:10.1038/s41699-020-0155-x.

13. Ermolaev, G.A.; Yakubovsky, D.I.; Stebunov, Y. V; Arsenin, A. V; Volkov, V.S. Spectral ellipsometry of monolayer transition metal dichalcogenides: Analysis of excitonic peaks in dispersion. *J. Vac. Sci. Technol. B* **2019**, *38*, 14002, doi:10.1116/1.5122683.

14. Li, H.; Contryman, A.W.; Qian, X.; Ardakani, S.M.; Gong, Y.; Wang, X.; Weisse, J.M.; Lee, C.H.; Zhao, J.; Ajayan, P.M.; et al. Optoelectronic crystal of artificial atoms in strain-textured molybdenum disulphide. *Nat. Commun.* **2015**, *6*, doi:10.1038/ncomms8381.

15. Cordovilla Leon, D.F.; Li, Z.; Jang, S.W.; Cheng, C.-H.; Deotare, P.B. Exciton transport in strained monolayer WSe2. *Appl. Phys. Lett.* **2018**, *113*, 252101, doi:10.1063/1.5063263.

16. Robert, C.; Amand, T.; Cadiz, F.; Lagarde, D.; Courtade, E.; Manca, M.; Taniguchi, T.; Watanabe, K.; Urbaszek, B.; Marie, X. Fine structure and lifetime of dark excitons in transition metal dichalcogenide monolayers. *Phys. Rev. B* **2017**, *96*, 155423, doi:10.1103/PhysRevB.96.155423.

17. Zhou, Y.; Scuri, G.; Wild, D.S.; High, A.A.; Dibos, A.; Jauregui, L.A.; Shu, C.; De Greve, K.; Pistunova, K.; Joe, A.Y.; et al. Probing dark excitons in atomically thin semiconductors via near-field coupling to surface plasmon polaritons. *Nat. Nanotechnol.* **2017**, *12*, 856–860, doi:10.1038/nnano.2017.106.

18. Klein, M.; Badada, B.H.; Binder, R.; Alfrey, A.; McKie, M.; Koehler, M.R.; Mandrus, D.G.; Taniguchi, T.; Watanabe, K.; LeRoy, B.J.; et al. 2D semiconductor nonlinear plasmonic modulators. *Nat. Commun.* **2019**, *10*, 3264, doi:10.1038/s41467-019-11186-w.

19. Camellini, A.; Mennucci, C.; Cinquanta, E.; Martella, C.; Mazzanti, A.; Lamperti, A.; Molle, A.; de Mongeot, F.B.; Della Valle, G.; Zavelani-Rossi, M. Ultrafast Anisotropic Exciton Dynamics in Nanopatterned MoS2 Sheets. *ACS Photonics* **2018**, *5*, 3363–3371, doi:10.1021/acsphotonics.8b00621.

20. Castellanos-Gomez, A.; Roldán, R.; Cappelluti, E.; Buscema, M.; Guinea, F.; van der Zant, H.S.J.; Steele, G.A. Local Strain Engineering in Atomically Thin MoS2. *Nano Lett.* **2013**, *13*, 5361–5366, doi:10.1021/nl402875m.

21. Quereda, J.; San-Jose, P.; Parente, V.; Vaquero-Garzon, L.; Molina-Mendoza, A.J.; Agraït, N.; Rubio-Bollinger, G.; Guinea, F.; Roldán, R.; Castellanos-Gomez, A. Strong Modulation of Optical Properties in Black Phosphorus through Strain-Engineered Rippling. *Nano Lett.* **2016**, *16*, 2931–2937, doi:10.1021/acs.nanolett.5b04670.

22. Wang, C.; Zhang, G.; Huang, S.; Xie, Y.; Yan, H. The Optical Properties and Plasmonics of Anisotropic 2D Materials. *Adv. Opt. Mater.* **2020**, *8*, 1–22, doi:10.1002/adom.201900996.

23. Li, L.; Kim, J.; Jin, C.; Ye, G.J.; Qiu, D.Y.; Da Jornada, F.H.; Shi, Z.; Chen, L.; Zhang, Z.; Yang, F.; et al. Direct observation of the layer-dependent electronic structure in phosphorene. *Nat. Nanotechnol.* **2017**, *12*, 21–25, doi:10.1038/nnano.2016.171.

24. He, Y.M.; Clark, G.; Schaibley, J.R.; He, Y.; Chen, M.C.; Wei, Y.J.; Ding, X.; Zhang, Q.; Yao, W.; Xu, X.; et al. Single quantum emitters in monolayer semiconductors. *Nat. Nanotechnol.* **2015**, *10*, 497–502, doi:10.1038/nnano.2015.75.

25. Mao, N.; Tang, J.; Xie, L.; Wu, J.; Han, B.; Lin, J.; Deng, S.; Ji, W.; Xu, H.; Liu, K.; et al. Optical Anisotropy of Black Phosphorus in the Visible Regime. *J. Am. Chem. Soc.* **2016**, *138*, 300–305, doi:10.1021/jacs.5b10685.

26. Castellanos-Gomez, A.; Agrat, N.; Rubio-Bollinger, G. Optical identification of atomically thin dichalcogenide

crystals. *Appl. Phys. Lett.* **2010**, *96*, doi:10.1063/1.3442495.

27. Li, Y.; Heinz, T.F. Two-dimensional models for the optical response of thin films. *2D Mater.* **2018**, *5*, doi:10.1088/2053-1583/aab0cf.

28. Mak, K.F.; Sfeir, M.Y.; Wu, Y.; Lui, C.H.; Misewich, J.A.; Heinz, T.F. Measurement of the Optical Conductivity of Graphene. *Phys. Rev. Lett.* **2008**, *101*, 196405, doi:10.1103/PhysRevLett.101.196405.

29. Liu, H.L.; Shen, C.C.; Su, S.H.; Hsu, C.L.; Li, M.Y.; Li, L.J. Optical properties of monolayer transition metal dichalcogenides probed by spectroscopic ellipsometry. *Appl. Phys. Lett.* **2014**, *105*, doi:10.1063/1.4901836.

30. Park, J.W.; So, H.S.; Kim, S.; Choi, S.H.; Lee, H.; Lee, J.; Lee, C.; Kim, Y. Optical properties of large-area ultrathin MoS2 films: Evolution from a single layer to multilayers. *J. Appl. Phys.* **2014**, *116*, doi:10.1063/1.4901464.

31. Rah, Y.; Jin, Y.; Kim, S.; Yu, K. Optical analysis of the refractive index and birefringence of hexagonal boron nitride from the visible to near-infrared. *Opt. Lett.* **2019**, *44*, 3797, doi:10.1364/ol.44.003797.

32. Schubert, M.; Rheinländer, B.; Franke, E.; Neumann, H.; Hahn, J.; Röder, M.; Richter, F. Anisotropy of boron nitride thin-film reflectivity spectra by generalized ellipsometry. *Appl. Phys. Lett.* **1997**, *70*, 1819–1821, doi:10.1063/1.118701.

33. Lee, S.Y.; Jeong, T.Y.; Jung, S.; Yee, K.J. Refractive Index Dispersion of Hexagonal Boron Nitride in the Visible and Near-Infrared. *Phys. Status Solidi Basic Res.* **2019**, *256*, 1–6, doi:10.1002/pssb.201800417.

34. DeFranzo, A.C.; Pazol, B.G. Index of refraction measurement on sapphire at low temperatures and visible wavelengths. *Appl. Opt.* **1993**, *32*, 2224, doi:10.1364/AO.32.002224.

35. Kuzmenko, A.B. Kramers-Kronig constrained variational analysis of optical spectra. *Rev. Sci. Instrum.* **2005**, *76*, 1–9, doi:10.1063/1.1979470.

36. Mayerhöfer, T.G.; Popp, J. Improving Poor Man's Kramers-Kronig analysis and Kramers-Kronig constrained variational analysis. *Spectrochim. Acta Part A Mol. Biomol. Spectrosc.* **2019**, *213*, 391–396, doi:10.1016/j.saa.2019.01.084.

37. Jackson, J.D. *Classical electrodynamics*; 3. ed., [N.; Wiley: Hoboken, NY, 2009; ISBN 978-0-471-30932-1.

38. Beal, A.R.; Liang, W.Y.; Hughes, H.P. Kramers-Kronig analysis of the reflectivity spectra of 3R-WS2 and 2H-WSe2. *J. Phys. C Solid State Phys.* **1976**, *9*, 2449–2457, doi:10.1088/0022-3719/9/12/027.

39. Beal, A.R.; Hughes, H.P. Kramers-Kronig analysis of the reflectivity spectra of 2H-MoS2, 2H-MoSe2 and 2H-MoTe2. *J. Phys. C Solid State Phys.* **1979**, *12*, 881–890, doi:10.1088/0022-3719/12/5/017.

40. Li, H.; Zhang, Q.; Yap, C.C.R.; Tay, B.K.; Edwin, T.H.T.; Olivier, A.; Baillargeat, D. From bulk to monolayer MoS 2: Evolution of Raman scattering. *Adv. Funct. Mater.* **2012**, *22*, 1385–1390, doi:10.1002/adfm.201102111.

41. Hecht, E. *Optics*; 3rd ed.; Addison-Wesley: Reading, Mass. :, 1998; ISBN 0201838877.

42. WebPlotDigitizer - Copyright 2010-2020 Ankit Rohatgi.

43. Yang, J.; Xu, R.; Pei, J.; Myint, Y.W.; Wang, F.; Wang, Z.; Zhang, S.; Yu, Z.; Lu, Y. Optical tuning of exciton and trion emissions in monolayer phosphorene. *Light Sci. Appl.* **2015**, *4*, e312--e312, doi:10.1038/lsa.2015.85.

44. Johnson, P.B.; Christy, R.W. Optical Constants of the Noble Metals. *Phys. Rev. B* **1972**, *6*, 4370–4379, doi:10.1103/PhysRevB.6.4370.

45. Ehrenreich, H.; Philipp, H.R. Optical Properties of Ag and Cu. *Phys. Rev.* **1962**, *128*, 1622–1629,

doi:10.1103/PhysRev.128.1622.

46. Cheng, F.; Su, P.-H.; Choi, J.; Gwo, S.; Li, X.; Shih, C.-K. Epitaxial Growth of Atomically Smooth Aluminum on Silicon and Its Intrinsic Optical Properties. *ACS Nano* **2016**, *10*, 9852–9860, doi:10.1021/acsnano.6b05556.

47. Maier, S.A. *Plasmonics fundamentals and applications*; Springer, /,: New York :, 2007; ISBN 9780387331508.

48. Bonifacio, L.D.; Lotsch, B. V; Puzzo, D.P.; Scotognella, F.; Ozin, G.A. Stacking the Nanochemistry Deck: Structural and Compositional Diversity in One-Dimensional Photonic Crystals. *Adv. Mater.* **2009**, *21*, 1641–1646, doi:10.1002/adma.200802348.

49. Paternò, G.M.; Manfredi, G.; Scotognella, F.; Lanzani, G. Distributed Bragg reflectors for the colorimetric detection of bacterial contaminants and pollutants for food quality control. *APL Photonics* **2020**, *5*, doi:10.1063/5.0013516.

50. Moscardi, L.; Paternò, G.M.; Chiasera, A.; Sorrentino, R.; Marangi, F.; Kriegel, I.; Lanzani, G.; Scotognella, F. Electro-responsivity in electrolyte-free and solution processed Bragg stacks. *J. Mater. Chem. C* **2020**, *8*, 13019–13024, doi:10.1039/d0tc02437f.

51. Normani, S.; Dalla Vedova, N.; Lanzani, G.; Scotognella, F.; Paternò, G.M. Design of 1D photonic crystals for colorimetric and ratiometric refractive index sensing. *Opt. Mater. X* **2020**, *8*, doi:10.1016/j.omx.2020.100058.

52. Ferry, V.E.; Munday, J.N.; Atwater, H.A. Design considerations for plasmonic photovoltaics. *Adv. Mater.* **2010**, *22*, 4794–4808, doi:10.1002/adma.201000488.